\documentclass[pre,aps,twocolumn]{revtex4}
\usepackage{amsfonts,amssymb,amsmath,bm}
\usepackage[dvips]{graphicx}
\usepackage{color}
\usepackage{times}
\usepackage{siunitx}

\begin{document}

\title{Effect of thin film on the generation of vorticity by surface waves}

\author{V.M. Parfenyev, S.S. Vergeles, and V.V. Lebedev}

\affiliation{
Landau Institute for Theoretical Physics RAS, Chernogolovka, 1-A Akademika Semenova av., 142432 Russia}

\date{\today}

\begin{abstract}
Recently a theoretical scheme explaining the vorticity generation by surface waves in liquids was developed [S. Filatov \textit{et al.}, Phys. Rev. Lett. {\bf 116}, 054501 (2016)]. Here we study how a thin (monomolecular) film presented at the surface of liquid affects the generated vorticity. We demonstrate that the vorticity becomes parametrically larger than for the case with a clean surface and now it depends on viscosity of the liquid. We also discuss the motion of particles passively advected by the generated surface flow. The results can be used in different applications: from the analysis of pollutants' diffusion on the ocean surface till the reconstruction of vorticity based on the particle image velocimetry (PIV) measurements.
\end{abstract}

\maketitle

\section{Introduction}
Boundary layers between different environments are crucial for biological, chemical and physical processes due to extreme conditions. For example, the sea surface is believed to play an important role in the origin and earlier evolution of life on the Earth \cite{biology}. The sea-surface microlayer is known to have a very complex structure, and sometimes the most upper layer is a monomolecular surface film formed by surfactant \cite{Soloviev}. Here we consider how such a thin film at a liquid surface modify hydrodynamic motion. It is well-known that the presence of a film increases the damping of surface waves \cite{Lamb, LL}. The history of this phenomenon dates back to antiquity, when ancient Greeks used oil to calm rough seas.
The effect is related to the film incompressibility that is correct for relatively slow surface waves. The films are formed by insoluble agents and therefore the film mass is conserved. That leads to the local conservation law for the film density and, as a consequence, to the additional hydrodynamic surface squeezing mode \cite{88KL,94KL}. The squeezing mode is faster than the gravitational-capillary waves, that justifies the incompressibility condition used at analyzing the damping of the gravitational-capillary waves.

Recently, we have developed a theoretical scheme explaining the vorticity generation by surface waves in liquids \cite{16FPVBLL}. The generation is explained by nonlinear hydrodynamic interaction. In the case where a thin film is present at the liquid surface the boundary conditions for the bulk motion are changed in comparison with the clean surface. Therefore the theoretical procedure has to be modified. In the paper we consider only films of negligible thickness (monomolecular), which can be formed, e.g., due to contamination of the ocean surface. We present the modified theoretical scheme concerning the vorticity generation, and compare the results to the case with a clean surface. In particular, we show that now the generated vorticity depends on kinematic viscosity $\nu$ of the liquid. The associated surface solenoidal velocity can be estimated as $v \sim \omega \kappa h^2$, where $\omega$ is the wave frequency, $h$ is the wave amplitude and $\kappa = \sqrt{\omega/\nu}$ is an inverse thickness of a viscous boundary sublayer. For the surface waves of the same amplitude the generated vorticity becomes parametrically larger than in the case with a clean surface. We demonstrate that this property is closely related to the increased damping of surface waves mentioned above. Some quantitative predictions, which can be checked experimentally, are formulated.

We also study a motion of light particles always staying at the surface and passively advected by the generated flow. In many experiments the velocity field at a liquid surface was obtained by the particle image velocimetry (PIV) method, e.g. \cite{16FPVBLL, 11KH, 14Shats}. Since the passive particles always stay at the liquid surface, they move not only horizontally, but also in a vertical direction with the surface. This vertical motion leads to the correction in measured velocity associated with the Stokes mechanism \cite{Stokes}. Analyzing the surface solenoidal currents one should take into account the effect upon treatment the experimental data. We speculate that the results concerning the motion of passive particles can also find applications in problems associated with spreading of pollutants on the ocean surface.

\section{Basic Equations}
We consider the bulk motion of a liquid, which obeys the Navier-Stokes equation \cite{LL,Lamb}
 \begin{eqnarray}
 \partial_t \bm v + (\bm v \nabla) \bm v
 = - \nabla P / \rho + \nu \nabla^2 \bm v,
 \label{NaSt}
 \end{eqnarray}
where $\rho$ and $\nu$ are the liquid mass density and the kinematic viscosity coefficient, respectively, $\bm v$ is the liquid velocity and $P$ is pressure. The equation (\ref{NaSt}) has to be supplemented by incompressibility condition \hbox{$\mathop{\mathrm{div}}{\bm v} = 0$}. The straightforward calculations lead to the equation for vorticity $\bm\varpi=\mathrm{curl}\, \bm v$
 \begin{equation}
 \partial_t \bm \varpi
 =-(\bm v \nabla) \bm \varpi
 +(\bm\varpi\nabla) \bm v
 +\nu \nabla^2 \bm \varpi.
 \label{vort1}
 \end{equation}

One should also supplement the Navier-Stokes equation (\ref{NaSt}) by the boundary conditions posed at the liquid surface. First, it is the kinematic boundary condition \cite{Lamb}
 \begin{equation}
 \label{kinem}
 \partial_t h = v_z - v_x \partial_x h - v_y \partial_y h,
 \end{equation}
implying that the liquid surface moves with the velocity $\bm v$. Here and thereafter we assume that the axis $Z$ is directed vertically, opposite to the gravitational acceleration $\bm g$ and that the equilibrium liquid surface coincides with plane $z=0$. The deviations from the equilibrium shape are described by the elevation $h(t,x,y)$, i.\,e. the liquid surface is determined by the equation $z = h(t,x,y)$. Note that the pressure $P$ in the Navier-Stokes equation (\ref{NaSt}) includes the gravitational term: $P=p+\rho g  z$, where $p$ is the internal pressure.

There is also the dynamic boundary condition that can be obtained from the requirement of zero momentum flux through the liquid surface. In the presence of a film on a liquid surface one has to take into account inhomogeneity of the surface tension coefficient $\sigma (t,x,y)$ related to its dependence on the film thickness. Therefore the boundary conditions at a liquid surface $z=h$ are modified in comparison with the free surface, see e.g. \cite{LL}, they are
 \begin{eqnarray}
 \label{DynBC_n}
 P - 2 \rho \nu l_i l_k \partial_i v_k
 =\rho g h + \sigma (\nabla \bm l),  \\
 \rho \nu \delta_{ij}^{\perp} l_k (\partial_j v_k + \partial_k v_j)
 =\delta_{ij}^{\perp} \partial_j \sigma.
 \label{DynBC_t}
 \end{eqnarray}
Here $\bm l (t,x,y) =  (-\partial_x h, -\partial_y h, 1)/\sqrt{g}$ is the unit vector normal to the surface, $g = 1+(\nabla h)^2$ can be thought as the determinant of the film metric tensor, and $\delta_{ij}^{\perp} = \delta_{ij} - l_i l_j$ is a projector operator on a film surface. The vorticity $\varpi_i = \epsilon_{ijk} \partial_j v_k$ should satisfy the boundary condition, which follows from Eq.~(\ref{DynBC_t})
 \begin{equation}
 l_m l_k \partial_k \varpi_m
 +(\partial_i v_k+\partial_k v_i)
 \epsilon_{imn} l_m K_{kn} = 0,
 \label{boundary2}
 \end{equation}
where $\epsilon_{ijk}$ is the unit antisymmetric tensor and we have introduced the curvature tensor $K_{ik} = K_{ki} = (\delta_{ij} - l_i l_j) \partial_j l_k$. Note that the gradient of the surface tension drops from the boundary condition (\ref{boundary2}).

To close the system of equations we need to know the dependence of surface tension $\sigma$ on the film density per unit area $n$. Moreover, since we have a new variable $n$, we should write down an additional equation. This is the mass conservation law for the film density:
\begin{equation}\label{boundary_gamma}
\partial_t (\sqrt{g} n) + \partial_{\alpha} (\sqrt{g} n v_{\alpha}) = 0,
\end{equation}
where the value velocity field should be taken at liquid surface $z=h$. Here and below Greek indices run over $x$ and $y$. The quantity $\sqrt{g} n$ is a projection of the film density on $X-Y$ plane. 
The equation (\ref{boundary_gamma}) can be rewritten as
 \begin{equation}
 (\partial_t +\bm v \nabla)\ln n +\delta^\perp_{ij}\partial_i v_j=0,
 \label{boundaryt}
 \end{equation}
where the relation (\ref{kinem}) was exploited.

\section{Linear Approximation}

Further we consider the case where some surface waves are excited in the liquid. The case of deep water is implied. We assume that the wave steepness is small, i.e. $|\nabla h|\ll1$. We also assume that the waves are weakly decaying, i.e. $\gamma = \sqrt{\nu k^2/\omega} \ll 1$, where $\omega$ is the wave frequency and $k$ is its wave vector.

At examining the hydrodynamic motion the film can be treated as incompressible in the linear approximation \cite{LL}, since the surface area of the film is changed only in the second-order in $|\nabla h|\ll1$. Therefore, we arrive at the surface incompressibility condition posed at $z=0$
 \begin{equation}
 \partial_{\alpha} v_{\alpha} = 0, \quad \partial_z v_z = 0,
 \label{surinc}
 \end{equation}
where we have used the three-dimensional incompressibility condition $\nabla \bm v=0$. Formally, the conditions (\ref{surinc}) can be obtained from Eq.~(\ref{boundaryt}) after neglecting nonlinear terms and the time derivative. Let us establish the corresponding criterion. It follows from Eq. (\ref{DynBC_t}) that $\delta\sigma\sim \rho\nu v$. Therefore $\delta n\sim (\partial\sigma/\partial n)^{-1} \rho\nu v$ and we arrive at the condition
 \begin{equation}
 \omega \rho\nu
 \ll k n\partial\sigma/\partial n.
 \label{criterion}
 \end{equation}

In the linear approximation, all quantities characterizing the surface waves can be expressed via the surface elevation $h$. The explicit expressions for velocity and vorticity in the leading order in parameter $\gamma$ are
\begin{eqnarray}
 \label{linear}
\displaystyle v_{\alpha} =  \nu \frac{\hat{\kappa}(\hat{\kappa}
 + \hat{k})}{\hat{k}} \left(e^{ \hat{k}z} -  e^{ \hat{\kappa} z} \right) \partial_{\alpha} h, &&\\
 \label{linear1}
 \displaystyle v_z = \nu (\hat{\kappa} + \hat{k}) \left( \hat{\kappa} e^{ \hat{k}z}
 - \hat{k} e^{ \hat{\kappa} z} \right) h, &&\\
 \label{linear2}
\varpi_{\alpha} = \epsilon_{\alpha \beta} \dfrac{\hat \kappa + \hat k}{\hat k} e^{\hat \kappa z} \partial_{\beta} \partial_t h + O(\gamma^2),&&
\end{eqnarray}
where we introduced nonlocal operators $\hat k=(-\partial_x^2-\partial_y^2)^{1/2}$, $\hat\kappa = (\partial_t/\nu + \hat{k}^2)^{1/2}$. The first terms in the right-hand sides of expressions (\ref{linear}), (\ref{linear1}) correspond to the potential part of velocity, whereas the last terms represent corrections, arising due to viscosity. The vorticity $\varpi_{\alpha}$ is located in a relatively thin layer near the surface. The depth of the layer is estimated as $\gamma/k \ll 1/k$, where $1/k$ is the penetration depth of the potential velocity.

The presence of film does not change the disrpersion law of surface waves, $\omega^2=gk +(\sigma_0/\rho) k^3$, except for the change in $\sigma_0$ --- an equilibrium value of surface tension $\sigma$. However, the wave damping is larger than in the clean case:
 \begin{equation}\label{damping}
 \frac{\mathrm{Im}\ \omega}{\omega}
 =\frac{\gamma}{2 \sqrt 2}.
 \end{equation}
Formally this happens because to satisfy the boundary condition (\ref{surinc}), the viscous contribution to the velocity field should be of the order of potential one, see Eqs.~(\ref{linear}) and (\ref{linear1}), while in the clean case the viscous contribution is smaller in parameter $\gamma$, see e.g. \cite{16FPVBLL}. Nevertheless, let us stress that the waves attenuate weakly due to the condition $\gamma\ll1$.

Note that the result (\ref{damping}) can be found in \cite{LL} for capillary waves. For these waves the criterion (\ref{criterion}) reads $\nu \sigma^{1/2} \rho^{1/2} k^{1/2} \ll  n\partial\sigma/\partial n$.

\section{Nonlinear Mechanism}

Principally, there is a contribution of the second-order in $\nabla h$ to $\delta n$. Then the first term in the boundary condition (\ref{boundaryt}) could be relevant for examining nonlinear effects. However, the contribution is irrelevant for the subsequent analysis, because the film density $n$ and the surface tension $\sigma$ do not enter in the boundary condition (\ref{boundary2}).

Thus, the generation mechanism of vertical vorticity is the same as in the clean case \cite{16FPVBLL}, except the change in the velocity field, see Eqs.~(\ref{linear}) and (\ref{linear1}). To find $Z$-component of the vorticity $\varpi_z$ we should solve the equation
 \begin{equation}
 (\partial_z^2-\hat\kappa^2)\varpi_z
 = - \nu^{-1}
 \varpi_\alpha \partial_\alpha v_z,
 \label{vort4}
 \end{equation}
supplemented by the boundary condition
 \begin{equation}
 \partial_z \varpi_z=
 \partial_\alpha h\partial_z \varpi_\alpha
 -\epsilon_{\alpha\gamma}
 (\partial_\alpha v_\beta+\partial_\beta v_\alpha)
 \partial_\beta \partial_\gamma h
 \label{boundary3}
 \end{equation}
posed at $z=0$ and $\varpi_z \rightarrow 0$ at $z \rightarrow -\infty$. The relation (\ref{boundary3}) is just the equation (\ref{boundary2}) written up to the second-order in $\nabla h$. Using Eqs.~(\ref{linear})-(\ref{linear2}), we find a solution
 \begin{equation}
 \begin{aligned}\label{vort_fin}
 \varpi_z (z) = \epsilon_{\alpha \beta} \left( e^{\hat \kappa z} \dfrac{\hat \kappa}{\hat k} \partial_{\beta} \partial_t h \right) \left( e^{\hat k z} \partial_{\alpha} h \right) + \\ \dfrac{\epsilon_{\alpha \beta}}{2}  \hat \kappa^{-1} e^{\hat \kappa z} \Big( (\partial_{\beta} \partial_t \hat k^{-1} h)(\hat \kappa \hat k \partial_{\alpha} h) - \dfrac{\hat \kappa}{\hat k} \partial_{\beta} h \partial_{\alpha} \partial_t \hat k h \Big).
 \end{aligned}
 \end{equation}
Here the first term represents the tilt of the vorticity (\ref{linear2}) due to the surface tilt and the other term is the result of rotated vorticity spreading into the bulk.

Since we consider the nonlinearity of the second-order, the characteristic frequency $\omega_v$ of vorticity $\varpi_z$ can vary from zero to the order of the surface wave frequency $\omega$. If $\omega_v \gg \nu k^2$ then the first term in the expression (\ref{vort_fin}) is leading, otherwise both terms are of the same order. Further we assume $\omega_v \ll \nu k^2$, then one can substitute $\hat\kappa$ by $\hat k$ in prefactor before brackets in the second line of Eq.~(\ref{vort_fin}). So, the first contribution in the right-hand side of Eq.~(\ref{vort_fin}) is localized on the scale $\gamma/k$ near the surface, while the second contribution penetrates deeper, on a distance $1/k$.

Let us consider the case of monochromatic pumping, when the absolute value of wave vector is fixed. The expression for the slow vorticity ($\omega_v \ll \nu k^2$) at a liquid surface can be simplified and it takes a form:
\begin{equation}\label{vort_fin_surf}
\varpi_z(0) = \epsilon_{\alpha \beta} \left( \dfrac{\hat \kappa}{\hat k} \partial_{\beta} \partial_t h \right) \partial_{\alpha} h +  \epsilon_{\alpha \beta} \hat k^{-1} (\hat \kappa \partial_{\alpha} h \partial_{\beta} \partial_t h ).
\end{equation}
To illustrate the relation we consider the case of two plane waves, propagating perpendicular to each other. Then the surface elevation can be modeled as
 \begin{equation}
 h=H_1 \cos(\omega t-kx)
 +H_2 \cos(\omega t-ky),
 \label{boundary36}
 \end{equation}
and substituting this expression to the equation (\ref{vort_fin_surf}), we obtain
\begin{equation}\label{answer}
\varpi_z (0) = - \dfrac{\sqrt{2}+1}{2 \gamma} H_1 H_2 \omega k^2 \sin(kx-ky).
\end{equation}

Note that the presented theory is correct if the higher-order nonlinear terms are small compared to the kept ones. We should estimate the nonlinear terms using Eq.~(\ref{vort1}), where the second-order terms for the velocity, $v^{(2)}$, have to be taken into account. From Eq.~(\ref{vort_fin_surf}) we find $v^{(2)}\sim\omega kh^2/\gamma$. Therefore the nonlinear terms with $v^{(2)}$ are small if $({\bm v}^{(2)}\nabla)\varpi_z \ll\nu\Delta\varpi_z$. Thus, in the case $\omega_\mathrm{v} \lesssim \nu k^2$ we arrive at the condition $kh\ll \gamma^{3/2}$, which is stronger than the weak steepness condition $kh\ll1$.

\section{Stokes Drift}

Now we analyze the motion of passive particles placed on a liquid surface and advected by the generated surface currents (\ref{vort_fin_surf}). Examining the trajectories of such particles is a natural way to observe and detect the generated vorticity. However, one should be careful upon treatment experimental data, because particles move not only horizontally, but also in a vertical direction with the surface. And this vertical motion lead to the correction in measured velocity field associated with the Stokes mechanism \cite{Stokes}.

The position of each particle can be characterized by a two-dimensional vector $\bm X = (X,Y)^{T}$, which obeys the equation of motion:
\begin{equation}\label{A1}
\dfrac{d \bm X}{dt} = \bm u (t, \bm X),
\end{equation}
where $\bm u (t, \bm X) \equiv \bm v (t,\bm X,h(t,\bm X))$ is horizontal velocity at a liquid surface. Near some point $\bm x_0 = (x_0, y_0)^{T}$ we can expand the velocity field in Taylor series:
\begin{equation}
\bm u (t, \bm x) = \bm u (t, \bm x_0) + \hat{G} \cdot \delta \bm x + \dots.
\end{equation}
Here $\delta \bm x = \bm x - \bm x_0$ and $\hat{G}$ is a velocity gradient tensor, its four components are
\begin{equation}
\begin{aligned}
G_{\alpha \beta} = \partial_{\alpha} u_{\beta} (t, \bm x_0) = \\ = \partial_{\alpha} v_{\beta} (t, \bm x_0, z) \vert_{z=h} + \partial_z v_{\beta} (t, \bm x_0, h) \partial_{\alpha} h.
\end{aligned}
\end{equation}
Now we solve the equation (\ref{A1}) up to the second-order in parameter $|\nabla h| \ll 1$ using iterative method. The particle displacement is $\delta \bm X = \delta {\bm X}_0 + \delta {\bm X}_1$, where
\begin{equation*}
\delta {\bm X}_0 = \int \bm u (t, \bm x_0) dt, \; \delta {\bm X}_1 = \int \hat{G} \cdot \delta {\bm X}_0 dt,
\end{equation*}
and we need to keep only linear terms in $\delta {\bm X}_0$ and $\hat{G}$ to calculate $\delta {\bm X}_1$.

Experimentally, the velocity field is reconstructed according to the definition $\delta \bm X / \delta t $. To find $\delta \bm X$ one should process consecutive images. If the time difference between these images is much smaller than the wave period $2 \pi/\omega$, we can neglect $\delta \bm X_1$ contribution and the velocity field captured by the fast camera is just $\bm u (t, \bm x)$. To obtain the stationary velocity we should average the expression over time (over many pairs of images).
In principle, it is also possible to use a slow camera for registration and capture only a slow motion. In this case the time difference between consecutive images must be a multiple of the wave period $2 \pi/\omega$, and we need to take $\delta \bm X_1$ contribution into account. For our particular case, due to the boundary condition (\ref{surinc}), one finds $\delta {\bm X}_1 = 0$ and therefore the reconstructed velocity field will be the same.

To calculate the vertical vorticity one should take curl from the reconstructed velocity, i.e. $\varpi_R = \langle \epsilon_{\alpha \beta} \partial_{\alpha} v_{\beta} (t, \bm x, h(t, \bm x)) \rangle$, where angular brackets denote averaging over time.  Up to the second-order, we obtain
\begin{equation}
v_{\beta} (t, \bm x, h) =  v_{\beta} (t, \bm x, 0) + h \partial_z v_{\beta} (t, \bm x, 0),
\end{equation}
and then
\begin{equation}
\varpi_R = \varpi_z (0) + \langle \epsilon_{\alpha \beta} \partial_{\alpha} h \partial_z v_{\beta} (t, \bm x, 0) \rangle.
\label{eq:stokes}
\end{equation}
Here we drop a term $h \partial_z \varpi_z (t, \bm x, 0)$, since it is of the third-order in $\nabla h$ according to the boundary condition (\ref{boundary3}). The last term in the expression (\ref{eq:stokes}) represents the correction to the previously obtained vorticity (\ref{vort_fin_surf}), associated with the Stokes mechanism. Finally, for the surface elevation given by the expression (\ref{boundary36}), we find:
\begin{equation}\label{answer2}
\varpi_R = - \dfrac{1}{2 \gamma} H_1 H_2 \omega k^2 \sin(kx-ky).
\end{equation}

\section{Discussion}

Comparing the results (\ref{vort_fin}) and (\ref{answer}) with ones obtained in Ref.~\cite{16FPVBLL}, we conclude that the answers are parametrically larger, they contain an additional factor $\gamma^{-1} \gg 1$. Thus, in the presence of the thin film the vertical vorticity generation by the surface waves is more effective than in the clean case (for the surface waves of the same amplitude). However, one should remember that in the presence of the film the surface waves are harder excited because of stronger damping.
In fact these two phenomena are closely related to each other. Indeed, according to the Eq. (\ref{vort4}) the source of vertical vorticity $\varpi_z$ is a horizontal vorticity $\varpi_{\alpha}$ slightly rotated by the velocity field. The vorticity $\varpi_{\alpha}$ is determined by non-potential contribution to the velocity field, and it contains the same additional factor $\gamma^{-1}$ due to the boundary condition (\ref{surinc}), when a thin film is presented at a liquid surface, see Eq. (\ref{linear2}) and Ref. \cite{16FPVBLL}. Such parametric increase in the horizontal vorticity leads to more effective generation of the vertical vorticity $\varpi_z$ and simultaneously to the stronger wave damping, see Eq. (\ref{damping}).
We also would like to note that exactly the same mechanism of vorticity generation takes place in a freely-suspended thin smectic films, which perform transverse oscillations \cite{16Parf}.

A natural way to observe the generated vorticity $\varpi_z$ is to examine trajectories of the passive particles placed on the liquid surface. Then one should take into account the Stokes mechanism \cite{Stokes}. It is interesting that the effect gives practically the same contribution to the particles' motion as the generated vorticity does, see Eq. (\ref{eq:stokes}). However, the Stokes mechanism contribution has an opposite sign and smaller amplitude than the direct effect of the generated vorticity. For the two perpendicular plane waves the resulting solenoidal motion of particles is described by the expression (\ref{answer2}).

Qualitatively, in the clean case situation is similar. For the specific examples of a liquid elevation $h(t,x,y)$ considered in the paper \cite{16FPVBLL}, the vorticity of Lagrangian-mean flow produced by the Stokes mechanism at a liquid surface (both discussed methods of velocity measurement also give the same result) has the same spatial structure as (14) and (16) in \cite{16FPVBLL}, but with the prefactor $2+\sqrt{2}$ replaced by $-1$. Thus, the resulting prefactor in the reconstructed vorticity $\varpi_R$ will be $1+\sqrt{2}$. Note, that there is an error in paper \cite{16FPVBLL} --- a color legends in Figs.~2 and 4 must be inverted. Taking into account this error we obtain that the direction of particles' motion in the Fig.~4 cannot be explained by the Stokes drift alone, since this mechanism moves particles in the opposite direction. Nevertheless, the quantitative comparison between theory and experiments is a necessary further step. We work in this direction, trying to take into account besides the Stokes drift also capillary effects \cite{Falkovich}. Results of the quantitative comparison will be published elsewhere.

\acknowledgments

We are grateful to O. B\"{u}hler for a comment about Stokes drift. This work was supported by the Russian Scientific Foundation, Grant No. 14-22-00259.

\end{document}